\documentclass[9pt]{IEEEtran}
\IEEEoverridecommandlockouts
\usepackage{amsmath,amssymb,amsfonts}
\usepackage[linesnumbered,ruled,noend]{algorithm2e}
\usepackage{graphicx}
\usepackage{textcomp}
\usepackage{xcolor}
\usepackage{booktabs}
\usepackage{multirow}
\usepackage{threeparttable}
\usepackage{listings}
\usepackage{amsmath}
\usepackage{subfigure}
\usepackage{flushend}
\usepackage{caption}
\usepackage{url}
\usepackage{placeins}
\usepackage[bottom]{footmisc}
\usepackage{textcomp}
\usepackage{float}
\usepackage{enumitem}
\usepackage[numbers,sort&compress]{natbib}
\def\BibTeX{{\rm B\kern-.05em{\sc i\kern-.025em b}\kern-.08em
    T\kern-.1667em\lower.7ex\hbox{E}\kern-.125emX}}

\definecolor{FDyellow}{RGB}{229,188,85} %
\definecolor{FDdark}{RGB}{27,67,150}   %
\definecolor{FDlight}{RGB}{115, 145, 192}  %
\definecolor{FDblue}{RGB}{46,91,161}    %
\definecolor{FDgreen}{RGB}{0,127,128}   %
\definecolor{FDorange}{RGB}{218,76,37} %
\definecolor{FDred}{RGB}{234,50,35}     %

\definecolor{citecolor}{RGB}{34,139,34}
\definecolor{mydarkblue}{rgb}{0,0.08,1}
\definecolor{mydarkgreen}{rgb}{0.02,0.6,0.02}
\definecolor{mydarkred}{rgb}{0.8,0.02,0.02}
\definecolor{mydarkorange}{rgb}{0.40,0.2,0.02}
\definecolor{mypurple}{RGB}{111,0,255}
\definecolor{myred}{rgb}{1.0,0.0,0.0}
\definecolor{mygold}{rgb}{0.75,0.6,0.12}
\definecolor{myblue}{rgb}{0,0.2,0.8}
\definecolor{mydarkgray}{rgb}{0.,0.2,0.2}

\definecolor{lightred}{RGB}{255,235,235}
\definecolor{lightgreen}{RGB}{235,255,235}
\definecolor{lightblue}{RGB}{235,235,255}
\definecolor{lightcyan}{RGB}{235,255,255}
\definecolor{lightmagenta}{RGB}{255,235,255}
\definecolor{lightyellow}{RGB}{255,255,235}

\definecolor{qxkcolor}{RGB}{215,235,255}
\definecolor{softmaxcolor}{RGB}{230,235,255}
\definecolor{probxvcolor}{RGB}{255,255,235}

\definecolor{topkcolor}{RGB}{255,235,235}
\definecolor{zecolor}{RGB}{255,255,235}
\definecolor{dynacolor}{RGB}{235,255,255}

\definecolor{reviewcolor}{RGB}{0,0,200}

\newcommand{\squishlist}{
 \begin{list}{$\bullet$}
  { \setlength{\itemsep}{0pt}
     \setlength{\parsep}{3pt}
     \setlength{\topsep}{3pt}
     \setlength{\partopsep}{0pt}
     \setlength{\leftmargin}{1.5em}
     \setlength{\labelwidth}{1em}
     \setlength{\labelsep}{0.5em} } }

\newcommand{\squishend}{
  \end{list}  }

\begin{document}
\pagestyle{empty}
\setlength{\textfloatsep}{5mm}
\title{Mixed Structural Choice Operator: Enhancing Technology Mapping with Heterogeneous Representations}

\author{\footnotesize\IEEEauthorblockN{Zhang Hu$^1$, Hongyang Pan$^2$, Yinshui Xia$^1$, Lunyao Wang$^1$ and Zhufei Chu$^1$\IEEEauthorrefmark{1}}

\IEEEauthorblockA{$^1$Faculty of Electrical Engineering and Computer Science, Ningbo University, Ningbo 315211, China}

\IEEEauthorblockA{$^2$School of Microelectronics, State Key Laboratory of Integrated Chips and System, Fudan University, Shanghai 200433, China}

\IEEEauthorblockA{\IEEEauthorrefmark{1}Email: chuzhufei@nbu.edu.cn}

}

\maketitle
\thispagestyle{empty}

\begin{abstract}

The independence of logic optimization and technology mapping poses a significant challenge in achieving high-quality synthesis results. Recent studies have improved optimization outcomes through collaborative optimization of multiple logic representations and have improved structural bias through structural choices. However, these methods still rely on technology-independent optimization and fail to truly resolve structural bias issues. 
This paper proposes a scalable and efficient framework based on \emph{\underline{M}ixed Structural \underline{Ch}oices} (MCH). This is a novel heterogeneous mapping method that combines multiple logic representations with technology-aware optimization.
MCH flexibly integrates different logic representations and stores candidates for various optimization strategies. By comprehensively evaluating the technology costs of these candidates, it enhances technology mapping and addresses structural bias issues in logic synthesis.

Notably, the MCH-based lookup table (LUT) mapping algorithm set new records in the \emph{EPFL Best Results Challenge} by combining the structural strengths of both And-Inverter Graph (AIG) and XOR-Majority Graph (XMG) logic representations. Additionally, MCH-based ASIC technology mapping achieves a 3.73\% area and 8.94\% delay reduction (balanced), 20.35\% delay reduction (delay-oriented), and 21.02\% area reduction (area-oriented), outperforming traditional structural choice methods. Furthermore, MCH-based logic optimization utilizes diverse structures to surpass local optima and achieve better results.

\end{abstract}


\section{Introduction}
Logic synthesis is one of the key steps in the \emph{electronic design automation} (EDA) workflow. The process of logic synthesis mainly includes translation, technology-independent logic optimization, and technology mapping, aiming to convert RTL designs into mapped netlists with optimized power, performance, and area (PPA). Achieving high-quality synthesis results is a significant and challenging scientific task.

During the logic synthesis phase, the design representation evolves from truth tables, \emph{sum of products} (SOPs), \emph{binary decision diagrams} (BDDs), to \emph{directed acyclic graphs} (DAGs), while the scale of the design gradually expands. Currently, mainstream logic representations in the logic synthesis process include \emph{AND-Inverter graph} (AIG), \emph{Majority-Inverter graph} (MIG)~\cite{MIG}, \emph{XOR-AND graph} (XAG)~\cite{XAG}, and \emph{XOR-Majority graph} (XMG)~\cite{XMG}. 
For different types of circuits, each representation has its unique advantages. For example, some novel logic representations perform better in arithmetic-intensive circuits~\cite{riener2019scalable}. In recent years, some studies have focused on exploring optimization methods and application for different logic representations~\cite{haaswijk2016lut,haaswijk2017novel,chu2019advanced,chu2019multi}. Modern IC designs consist of complex functional modules that can be specifically optimized using different logic representations.

Recent studies have achieved better optimization results by combining different logic representations~\cite{amaru2013mixsyn,neto2019lsoracle,austin2020scalable,neto2022flowtune}. For example, the LSOracle framework~\cite{neto2019lsoracle,austin2020scalable} introduces a heterogeneous synthesis approach. It partitions the circuit into segments optimized using different logic representations, which are later unified into a single representation for mapping. However, these methods are often limited by \textbf{\emph{structural bias}}, a well-known issue in logic synthesis~\cite{chatterjee2006reducing}. They convert all representations into a single form for mapping, which compromises the structural advantages of other representations.

\begin{figure}[t]
    \centering
    \subfigure[Mapping area information]{
        \includegraphics[width=0.23\textwidth]{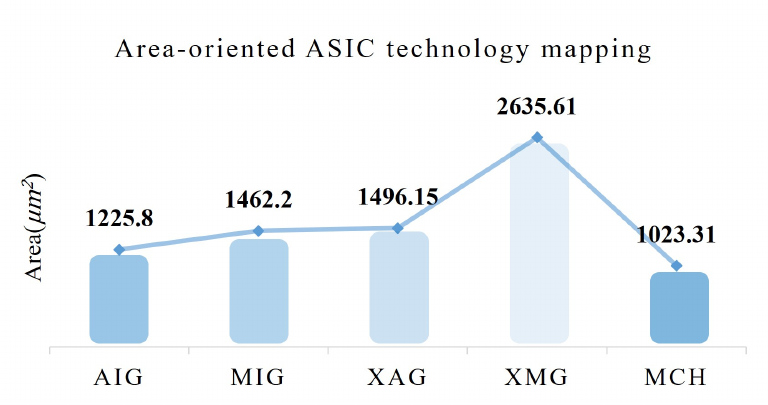}
    }\hspace{-0.3em}
    \subfigure[Mapping delay information]{
        \includegraphics[width=0.23\textwidth]{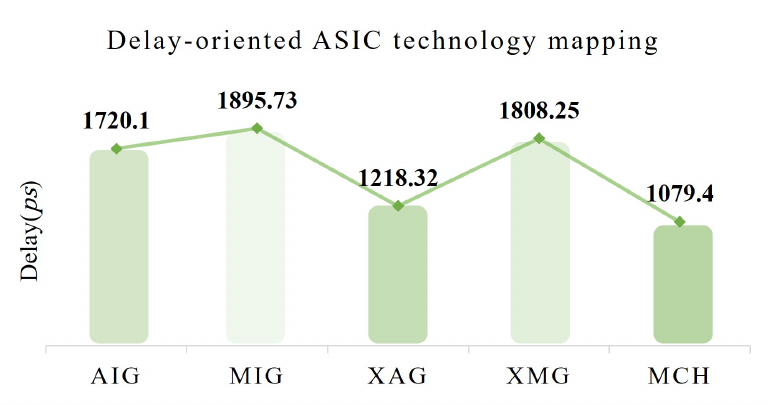}
    }
    \caption{Technology mapping for different logic representations}
    \label{map case}
    \vspace{-4mm}
\end{figure}

As design types and requirements become increasingly complex, the limitations of relying on a single logic representation are becoming more apparent. Fig.~\ref{map case} illustrates the conversion of the ``Max" circuit from the \emph{EPFL benchmark suite}~\cite{amaru2015epfl} into various logic representations, along with information on technology mapping using the \emph{Arizona State Predictive PDK 7nm} (ASAP7) library~\cite{clark2016asap7}. Results indicate that, for delay-oriented mapping, using XAG yields better performance for this circuit, while for area-oriented mapping, AIG achieves superior results. The mapping results are strongly influenced by the circuit's structure. Therefore, an effective technology mapping method should be able to jointly evaluate different logic representations, enhancing adaptability across different synthesis scenarios to achieve superior mapped netlists. In fact, our method achieves this by utilizing \emph{Mixed Structural Choices} (MCH) to effectively overcome the \emph{structural limitations} of single logic representations, resulting in more competitive outcomes.

The issue of structural bias permeates all stages of logic synthesis. The root cause lies in the independence between logic optimization and technology mapping. This causes logic optimization may misalign with mapping, wasting resources and missing optimal structures. Some studies employ machine learning and heuristic approaches to explore better mapping structures~\cite{neto2021slap,wang2023easymap,liu2023aimap}. Another promising approach is structural choices, which preserve different structural candidates during logic optimization to address structural bias~\cite{mishchenko2005technology,chatterjee2006reducing,mishchenko2010global,grosnit2023lightweight}. However, these methods are limited to searching for optimal mapping structures within the optimized graph. It is difficult to cope with the structural limitations of the graph when the optimization is close to a local optimum.
On the other hand, existing structural choice methods rely on equivalence checks between technology-independent optimized networks, significantly limiting the diversity of candidates and reducing the efficiency of constructing choice networks.
In recent studies, E-Syn utilize e-graphs to explore optimal structures~\cite{chen2024syn}. However, its efficiency is relatively low when handling large-scale designs.

\begin{figure}[t]
\centering
\includegraphics[scale=0.32]{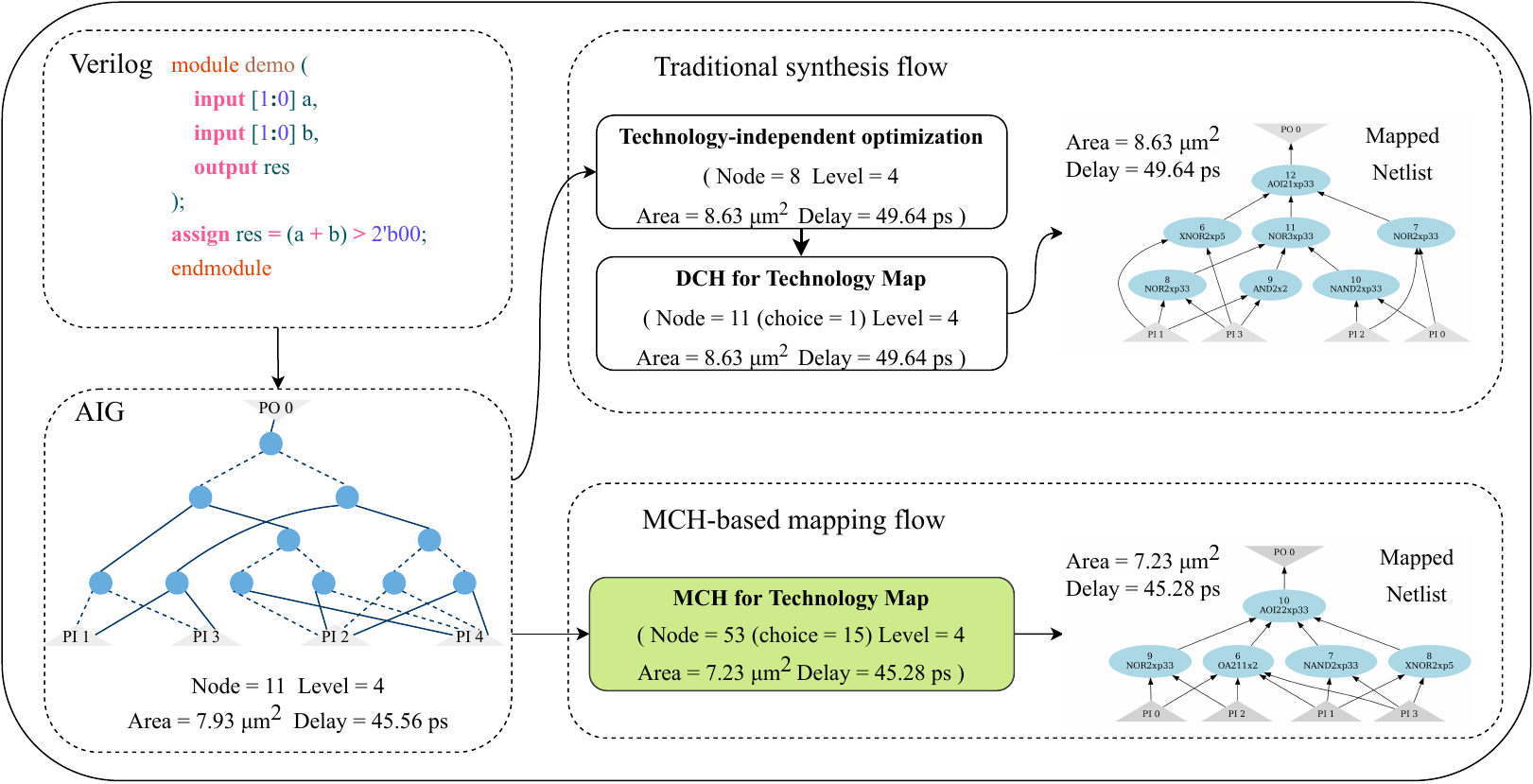}
\caption{Comparison of the MCH-based mapping flow with the traditional synthesis flow}
\label{structural bias}
\vspace{-4mm}
\end{figure}

Fig.~\ref{structural bias} illustrates the limitations of the traditional synthesis flow through a simple example. The original network is an AIG with 11 nodes. After technology-independent optimization, the number of AIG nodes is reduced to 8, but the mapping cost increases instead. Subsequently, traditional structural choice algorithms~\cite{chatterjee2006reducing} (denoted as DCH in ABC~\cite{ABC}) still fail to effectively address the impact of structural limitations. This issue is particularly common in large-scale designs. Thus, achieving synergy between logic optimization and technology mapping remains a significant technical challenge.

Based on existing issues, this paper introduces the \emph{Mixed Structural Choices} (MCH) operator. They are mixed networks incorporating heterogeneous logic representations, eliminating the constraints of relying on a single logic representation during logic synthesis.
Unlike existing heterogeneous synthesis methods, MCH preserves the original logic structure through structural choices while incorporating new logic representations as candidates.
Furthermore, to obtain diversified candidate structures, MCH introduces a multi-strategy structural choices algorithm. It combines node path classification with synthesis strategies targeting different objectives, preserving equivalence rather than performing local replacements before optimization, thereby retaining diverse optimization candidates.
Finally, MCH integrates with technology mapping, using technology information to evaluate candidate logic representations and structures, thereby enhancing technology mapping.

The main contributions of this paper are as follows:  
\begin{itemize}
\item[$\bullet$]The concept of MCH is proposed, enabling the simultaneous evaluation of different logic representations through choice networks.
\item[$\bullet$]Multi-strategy structural choices differentiate critical paths and apply diverse optimization strategies, preserving varied logic representations and optimization candidates.
\item[$\bullet$]Forms an MCH-based technology mapping method, which evaluates candidate logic representations and optimizations using technology information.
\item[$\bullet$]In tests based on the EPFL benchmark suite, MCH-based \emph{application specific integrated circuits} (ASIC) technology mapping achieved improvements across multiple metrics. MCH-based \emph{field programmable gate array} (FPGA) technology mapping set new records in the \emph{EPFL Best Results Challenge}. MCH-based graph mapping overcomes local optima, improving logic optimization and post-mapping results. 
\end{itemize}

\section{Background}
\subsection{Boolean Network}

In logic synthesis, Boolean networks are typically represented as directed acyclic graphs. DAGs consist of logic nodes representing Boolean functions, \emph{primary inputs} (PI), \emph{primary outputs} (PO), and directed edges connecting these nodes. The level of node \( n \) is defined as the length of the longest path from any PI to node \( n \). The level of a Boolean network is determined by the maximum level of its internal nodes.
The output of a node \( n \) can connect to other nodes, known as the node's fanout, and the set of all fanout nodes forms the \emph{transitive fanout} (TFO) cone. The \emph{transitive fanin} (TFI) cone of a node \( n \) includes all nodes reachable through its fanin connections. A cut \( C \) for a node \( n \) in a DAG is a set of nodes that represent a portion of the logic function, where the nodes in \( C \), known as leaf nodes, ensure that every path from the PI to \( n \) passes through at least one leaf node. The cut enumeration algorithm ~\cite{mishchenko2007combinational} can compute the set of cuts for all nodes in a Boolean network for further analysis. The \emph{maximum fanout-free cone} (MFFC) of a node \( n \) is a subset of its TFI, where all paths to any PO pass through \( n \).

\subsection{Technology Mapping}

Technology mapping is the process of converting a logic network represented as a DAG into a netlist composed of actual technology library cells. Technology mapping is typically divided into standard cell mapping for ASICs and \emph{K-input lookup table} (K-LUT) mapping for FPGAs. This process involves steps such as cut enumeration, Boolean matching, and iterative mapping, using gates from the library that implement the same Boolean functions to cover the nodes in the network. The quality of technology mapping is usually evaluated in terms of area, delay, and power consumption. A delay-oriented mapping focuses on reducing the delay along the longest path in the cover, while an area-oriented mapping seeks to minimize the total area of the cover~\cite{calvino2022versatile}.

The quality of the mapped netlist depends on the matching possibility between the technology library cells and the DAGs to be mapped. Therefore, the quality of the mapped netlist is influenced by the DAGs to be mapped, and two DAGs representing the same Boolean function but with different structures may lead to significantly different results~\cite{lehman1997logic}.

\subsection{Structural Choices}
\begin{figure}[htbp]
\centering
\includegraphics[scale=0.6]{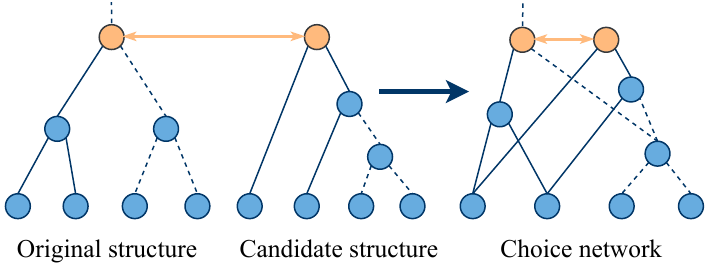}
\caption{Combine networks to generate a choice network}
\label{choice network}
\vspace{-4mm}
\end{figure}

Structural choice networks retain different structures that are functionally equivalent or complementary within a Boolean network. Fig.~\ref{choice network} illustrates an example of a choice network. A cut-based technology mapper can consider different options to decide which of the two equivalent structures to cover, which often helps alleviate structural bias, resulting in higher-quality outcomes~\cite{mishchenko2006improvements}. 
The concept of lossless logic synthesis was introduced in~\cite{chatterjee2006reducing}, applying choice networks to address structural bias issues. However, the creation of choice networks relies on technology-independent optimization, leading to structural limitations. 
LCH effectively addresses the inefficiency of traditional structural choice algorithms but does not resolve the inherent structural limitations~\cite{grosnit2023lightweight}.
A novel structural choice method based on logic optimization candidates is used to address structural bias issues; however, its support for large-scale circuits remains limited~\cite{hu2024novel}.

\section{OUR APPROACH}
\begin{figure*}[htpb]
\centering
\includegraphics[scale=0.45]{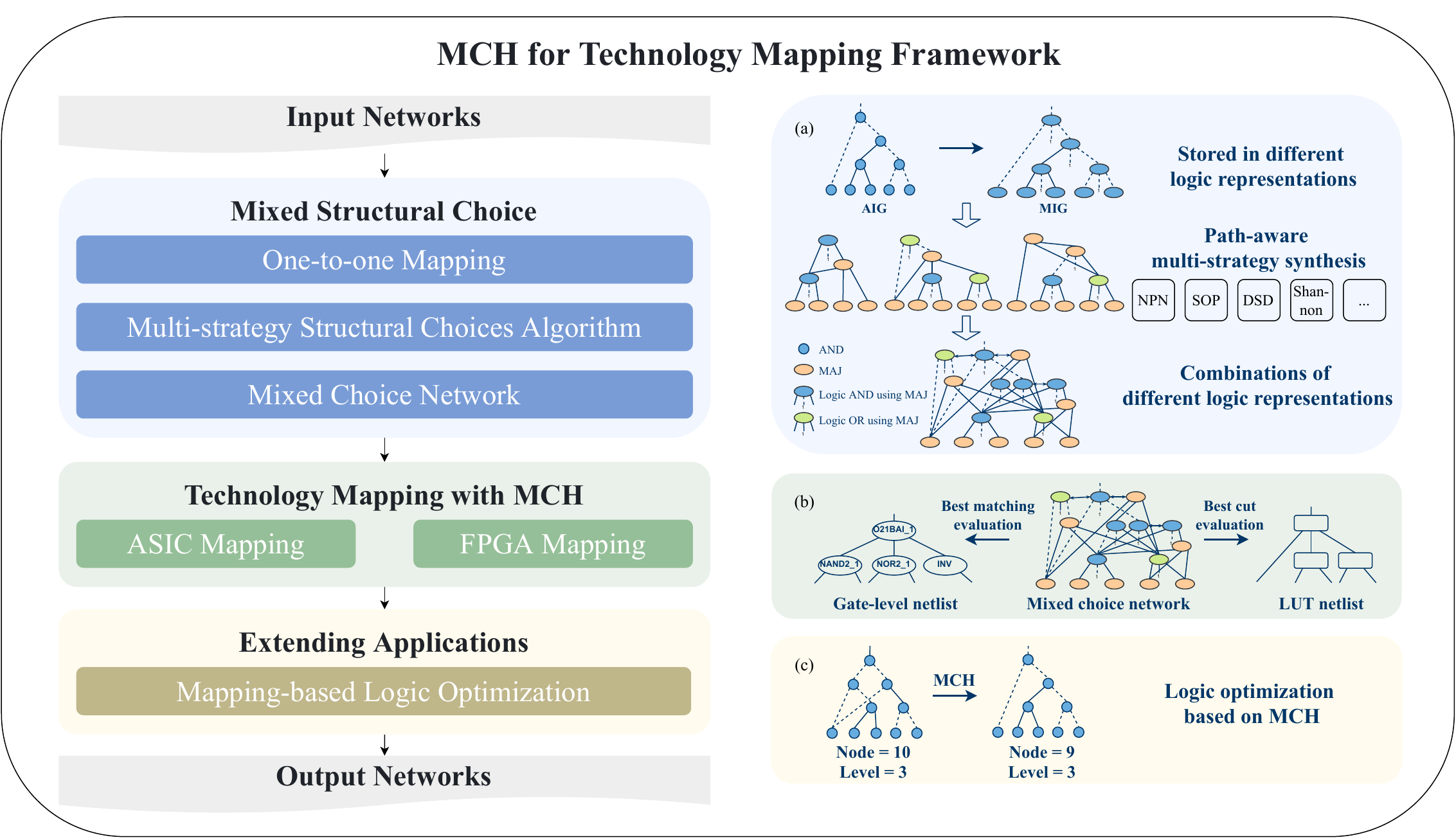}
\caption{Mixed Structural Choices for Technology Mapping Framework}
\label{framework}
\end{figure*}

This chapter provides an overview of the workflow for the \emph{Mixed Structural Choices for Technology Mapping} framework. The framework is a heterogeneous mapping approach that integrates mixed choice networks, technology-aware evaluation, and mapping-base optimization methods. 
As shown in the Fig.~\ref{framework}, the main innovations of this work consist of three parts: 
\begin{enumerate}[label=(\alph*)]
    \item \emph{Mixed Structural Choice}: This section explains the concept and creation of mixed structural networks. It preserves the structural characteristics of different logic representations through mixed structural networks and obtains candidate structures based on a path-classification multi-strategy synthesis approach.
    \item \emph{Technology Mapping with MCH}: This section introduces the technology mapping algorithm based on MCH, which evaluates candidates in MCH using technology mapping costs to enhance the mapping process.
    \item \emph{Extending Applications}: This section presents the application of MCH in mapping-based logic optimization, demonstrating its excellent scalability.
\end{enumerate}

\subsection{Mixed Structural Choices}

This section introduces the mixed structural choices, a heterogeneous logic representation approach based on structural choices, capable of integrating structural characteristics from different logic representations. MCH retains the original structure while storing other logic representations as candidates, enabling simultaneous evaluation of different logic representations during mapping. This enables the technology mapping phase to select the most suitable logic form, yielding higher-quality mapping results. 

AIG, as one of the most popular logic representations, is structurally compatible with XAG, MIG, and XMG. Therefore, we can create mixed choice networks based on AIG combined with any logic representation.
Fig.~\ref{framework} (a) illustrates an example of an mixed choice network, where the original network is an AIG network with a simple structure, while its choice nodes are in MIG, a form with more compact logic expressions. Through MCH, both structural characteristics are simultaneously available, allowing technology mapping to overcome the limitations of single logic representations and achieve an optimized logic network structure with enhanced mapping quality.

The MCH construction algorithm flow is shown in Algorithm~\ref{algo1}. 
The algorithm takes a Boolean network $\mathbb{N}$, a maximum cut size parameter $k$, a maximum number of cuts for each node parameter $l$, a maximum number of PIs in the MFFC parameter $K$, a constant parameter $r$ that controls the proportion of critical path nodes and the synthesis strategies library $lib$ as inputs. The output of the algorithm is the mixed choice network $G$.

\setlength{\textfloatsep}{2mm}
\begin{algorithm}[b]
\small
\caption{Mixed Structural Choices}
\label{algo1}
\BlankLine
 \KwIn{Boolean network $\mathbb{N}$, cut size $k$, cut limit $l$, MFFC max\_pi $K$, ratio $r$, synthesis strategies library $lib$}
    \KwOut{Mixed choice network $G$}
    more expressive logic representation $\mathbb{N'}$ $\gets$ \texttt{One-to-One\_Mapping}($\mathbb{N}$);\par
    hash table $f$ $\gets$ \texttt{Critical\_Path\_Collection}($\mathbb{N'}$, $r$);\par
    cuts set $C$ $\gets$ cut enumeration ($\mathbb{N'}$, $k$, $l$);\par
    \texttt{MCH} \emph{info.} $\gets$ \texttt{Multi-strategy\_Structural\_Choices} ($\mathbb{N'}$, $f$, $C$, $K$, $lib$); \hfill // Algorithm 2\par 
        $G$ $\gets$ generate structural choice network (\emph{info.});\par
        return $G$;
\end{algorithm}

When creating MCH, the first step involves storing the input network within a different logic representation format (line 1). For instance, the input AIG network is one-to-one mapped into the MIG data structure. This process traverses the input AIG nodes and creates corresponding AND nodes, constants, PIs, and POs within the MIG structure, ensuring that the original AIG characteristics are fully retained. 

Subsequently, MCH uses a hash table to collect critical path nodes, forming the critical path node set $f$. These nodes include PO nodes with logic depths greater than or equal to the network logic depth * $r$, along with all nodes on paths from these POs to the PIs (line 2). Next, MCH runs a cut enumeration algorithm to calculate cut information, which is saved according to cut size $k$ and cut number limit $l$ (line 3).

After completing the above steps, MCH uses a \emph{multi-strategy structural choice algorithm} to obtain diversified candidate information, as shown in Algorithm~\ref{algo2}. \emph{Mixed structural choices information} (\emph{info.}) refers to the equivalence information between the obtained logic representations and optimization candidates and the original network nodes. The algorithm traverses all nodes. For nodes in the critical path node set $f$, the focus is on obtaining structures with level advantages. Therefore, the cuts for these nodes are processed using level-oriented synthesis strategies, such as the 4-input NPN library~\cite{NPN}, to generate a diverse set of candidate options (lines 2-6). For non-critical path nodes, their MFFCs are computed (line 8). Area-oriented synthesis strategies, such as SOP and DSD, are then applied to generate candidate options with area advantages (lines 9-13). It is worth noting that these synthesized subcircuits do not replace the original network but are retained alongside it as structurally different but functionally equivalent options. After synthesis and restructuring, the obtained candidates will be preserved with new logic representation structures. The \emph{multi-strategy structural choice algorithm} further enriches the diversity of structural choices by combining path classification with multi-objective logic optimization strategies.

\setlength{\textfloatsep}{2mm}
\begin{algorithm}[t]
\small
\caption{Multi-strategy Structural Choices Algorithm}
\label{algo2}
\BlankLine
 \KwIn{Boolean network $\mathbb{N'}$, hash table $f$, cuts set $C$, MFFC max\_pi $K$, synthesis strategies library $lib$}
    \KwOut{Mixed Structural Choices information \emph{info.}}
        \For{each gate $n$ $\in$ $\mathbb{N'}$}{
        \If{$n \in f$}{
        \For{each cut $c$ $\in$ $C(n)$}{
        candidate set $E$ $\gets$ Level-oriented synthesis ($lib$, $c$);\par
        }
        \For{each gate $n'$ $\in$ $E$}{
        \emph{info.} $\gets$ mark representative and choice nodes;\par
        }
        }
        \Else
        {MFFC $m$ $\gets$ MFFC computation ($\mathbb{N'}$, $K$);\par
        \For{each cut $c$ $\in$ $C(n)$}{
        candidate set $E$ $\gets$ Area-oriented synthesis ($lib$, $c$);\par
        }
        candidate set $E$ $\gets$ Area-oriented synthesis ($lib$, $m$);\par
        \For{each gate $n'$ $\in$ $E$}{
        \emph{info.} $\gets$ mark representative and choice nodes;\par
        }
        }
        }
        return \emph{info.};
\end{algorithm}

Finally, the mixed choice network $G$ is constructed based on MCH information (line 5). Nodes in the original network are defined as \texttt{representative nodes}, and their equivalent candidates are added to the network as \texttt{choice nodes}. This MCH creation approach achieves a balance between depth and area, and MCH can also create networks optimized for area or delay by adjusting parameter $r$ and synthesis strategies to meet different design goals.

\subsection{Technology Mapping with MCH}

Thanks to the characteristics of MCH, we can achieve synergy between logic optimization and technology mapping. This section will explore how to combine MCH with technology mapping algorithms to utilize technology information for technology-aware evaluation. 

MCH-based technology mapping integrates the evaluation and selection of candidate structures from the choice network into existing mapping steps, guiding the mapper to choose the optimal structure for mapping.
The process pseudocode for the MCH-based technology mapping algorithm is shown in Algorithm~\ref{algo3}. The input of this algorithm is the mixed choice network $G$ obtained from subsection A, and the output is the mapping netlist $\mathbb{M}$. Thanks to the high flexibility of MCH, our method supports technology mapping for both ASIC and FPGA. 

First, the mixed choice network is traversed to compute and save each node's cuts (line 1). Next, MCH transfers the candidate structural information from choice nodes to their representative nodes. This process involves iterating through the cut sets of choice nodes and integrating their cuts into the cut sets of the corresponding representative nodes. Finally, a simple sorting is performed to remove cuts exceeding the cut set limit (lines 2-8). When the graph to be mapped contains some structurally different nodes marked as equivalent, cut-based technology mappers need to evaluate different covering options to select the optimal structure for coverage. Therefore, sharing the cut sets of choice nodes is essential.

During technology mapping, the mapper sorts the cut sets based on mapping objectives, selects the optimal cut, and determines the best cell by combining required time and Boolean matching. Since the candidate structures for logic representation and optimization have been successfully stored in the mixed choice networks, the mapper can select the structure with the lowest technology cost based on mapping requirements and technology information. Mapping is a dynamic programming process, for ASIC mapping, the MCH-based mapper continuously updates and selects the cut and matching cell with the lowest cost from numerous candidates. For FPGA mapping, the best cut is dynamically selected in each iteration based on the mapping target (lines 9-13).
Finally, the mapped netlist is generated and output, completing the mapping process (lines 14-15).

MCH-based technology mapping achieves a seamless integration of logic optimization and technology mapping. The heterogeneous logic representations and optimization candidates stored in MCH are directly utilized by the mapper, which selects structures with lower technology costs from the candidates. This approach addresses the limitations of the original network structure and directly improves the QoR of technology mapping.

\setlength{\textfloatsep}{2mm}
\begin{algorithm}[t]\small
    \caption{MCH-based technology mapping}\label{algo3}
    \BlankLine
    \KwIn{Mixed choice network $G$, cut limit $l$, tech library $tech\_lib$}
    \KwOut{Mapping netlist $\mathbb{M}$}
    cuts set $C$ $\gets$ Cut enumeration ($G$, $l$);\par
    \For{each gate $n$ $\in$ $G$} {
    $n'$ $\gets$ $n$'s choice node;\par
    \While{$n'$ have choice node}{
        \For{each cut $c$ $\in$ $C(n')$}{   
        insert cut $c$ to $C(n)$;\par
        }
        $n'$ $\gets$ $n'$'s choice node;\par
        }
        limit $C(n)$'s size $<=$ $l$;\par
    }
    \For{compute mapping} {
    \For{each gate $n$ $\in$ $G$} {
    sort $C(n)$ (delay / area);\par
    compute required time;\par
    compute delay-oriented / area-oriented mapping ($G$, $C$, $tech\_lib$);\par
    }
    }
    $\mathbb{M}$ $\gets$ generate mapping netlist;\par
    return $\mathbb{M}$;
\end{algorithm}

\subsection{Extending Applications}
Structural bias partly arises from the limitations of logic optimization, which are related to the logic optimization operators applied. Due to constraints imposed by graph structures or the optimization algorithms themselves, logic optimization algorithms often fall into local optima, thereby limiting the discovery of global optimal solutions. This phenomenon has been confirmed in recent \emph{IWLS Programming Contests}\footnote{\url{https://www.iwls.org/iwls2024/}}, where many participating teams have focused on addressing this issue.

MCH offers a new approach to addressing such issues. MCH can be easily extended to mapping-based logic optimization algorithms. MCH-based logic optimization leverages its unique features, combining different logic representations to help the original algorithms overcome local optima. MCH inherently provides a broader range of optimization candidates, and when integrated with logic optimization, it further expands the solution space, making it possible to achieve better optimization results.

\begin{figure}[htpb]
\centering
\includegraphics[scale=0.7]{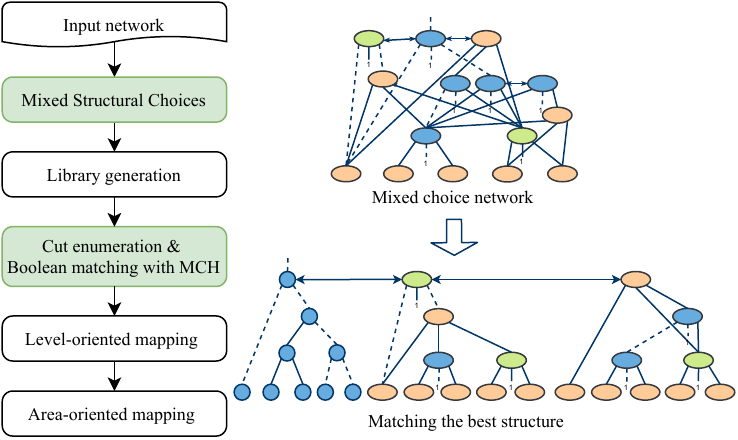}
\caption{MCH-based graph mapping}
\label{Extending_Applications}
\vspace{-4mm}
\end{figure}

\emph{Graph mapping} is a mapping-based method that remaps the original network to achieve the conversion between logic representations or to optimize logic networks~\cite{calvino2022versatile}. Fig.~\ref{Extending_Applications} illustrates the MCH-based graph mapping algorithm. The newly added work is marked in green in the figure. Our method uses MCH as an intermediate representation for graph mapping, performing cut enumeration and Boolean matching based on the mixed choice network. As shown in the figure, the algorithm can select the most advantageous structure for matching and mapping from three candidate structures. It then completes subsequent mapping and ultimately produces an optimized logic network. At this point, the output network corresponds to the new logic representation selected from the MCH candidates. Therefore, when implementing logic optimization, the output network can be mapped one-to-one to the target logic representation again. And then MCH-based graph mapping is performed to obtain the final optimized network. Through iterative cycles, the structural characteristics of multiple logic representations are combined to achieve higher-quality graph mapping optimization.

\section{Experimental Results}
In this section, we conducted experiments and evaluations of the technology mapping and logic optimization based on MCH. The experiments include technology mapping based on MCH, an evaluation of structural choice effects (comparing MCH and DCH), and an assessment of logic optimization results with MCH. We developed algorithms related to MCH based on the open-source logic synthesis framework \emph{Mockturtle}~\cite{soeken2018epfl} and conducted experiments. All experiments were performed on a 3.70 GHz AMD(R) Ryzen 5(R) 7500F CPU with 32GB of main memory. Our test cases come from the EPFL benchmark suite~\cite{amaru2015epfl} which includes arithmetic and random\_control benchmark circuits. All results have been formally verified with ABC's \texttt{cec} command.

\subsection{MCH for Technology Mapping}

\subsubsection{ASIC Technology Mapping}

In the ASIC technology mapping experiment, we evaluated the effectiveness of technology mapping based on MCH. The experiment first used ABC's \texttt{compress2rs} flow for iterative optimization to simulate the logic optimization process. Then, the mapping quality is evaluated using the ASAP7 technology library. The experimental results are presented in Table \ref{ASIC mapping}, which summarizes the area ($\mu\text{m}^2$) and delay ($ps$) information reported from the mapped netlist. The CPU runtime (Time) is recorded in seconds. The geometric mean (Geomean) and gain (Improvements) of each data point are also presented. 

The experiment primarily compares mapping and structural choices algorithms in ABC. We use the netlist results obtained from the command combinations for different mapping objectives in ABC as the baseline. In contrast, \emph{MCH Balanced} creates the choice network based solely on the input AIG, with candidate structures generated through a strategy that combines critical path and non-critical path node classification. Finally, achieving a balanced improvement in the mapping results. \emph{MCH Delay-oriented} adopts a more aggressive strategy. First, it transforms the input AIG into an XAG. Then, it retains the structural advantages of XAG and creates mixed choice networks of XAG and AIG. During the creation of candidate structures, the range of collected critical path nodes is also expanded. Finally, a delay-oriented mapping is performed. \emph{MCH Area-oriented} combines AIG and XMG to create mixed choice networks and performs area-oriented mapping based on MCH. 

The results show that \emph{MCH Balanced} achieves a 3.73\% area reduction and a 8.94\% delay optimization compared to the baseline. In the experiments for delay-oriented and area-oriented mapping scenarios, DCH yields very limited optimization benefits due to structural bias. In comparison, MCH leverages the structural advantages of XAG and XMG, achieving superior mapping results. Specifically, \emph{MCH Delay-oriented} achieves a 20.35\% delay improvement while sacrificing only 9.75\% of the area. On the other hand, \emph{MCH Area-oriented} achieves a similar area gain to ABC but significantly reduces the delay. This is because MCH utilizes heterogeneous representations to offer more possibilities for mapping. Additionally, although the computation and evaluation of candidate structures slightly increase mapping time. But for the larger circuits such as ``Hypotenuse", MCH can achieve more competitive results in a shorter time. This highlights the significant advantage of MCH for large-scale designs.

\begin{table*}[htbp]
  \centering
  \caption{Experimental results of ASIC technology mapping}
  \setlength{\tabcolsep}{0.65mm}
  \scriptsize
    \begin{tabular}{lrrrrrrrrrrrrrrrrrrrrrrr}
    \toprule
    \multicolumn{1}{c}{\multirow{2}[3]{*}{Benchmarks}} & \multicolumn{3}{c}{\&nf} &       & \multicolumn{3}{c}{\&dch -m;\&nf} &       & \multicolumn{3}{c}{dch;map -a} &       & \multicolumn{3}{c}{MCH balanced} &       & \multicolumn{3}{c}{MCH Delay-oriented} &       & \multicolumn{3}{c}{MCH Area-oriented} \\
\cmidrule{2-4}\cmidrule{6-8}\cmidrule{10-12}\cmidrule{14-16}\cmidrule{18-20}\cmidrule{22-24}          & \multicolumn{1}{c}{Area} & \multicolumn{1}{c}{Delay} & \multicolumn{1}{c}{Time} &       & \multicolumn{1}{c}{Area} & \multicolumn{1}{c}{Delay} & \multicolumn{1}{c}{Time} &       & \multicolumn{1}{c}{Area} & \multicolumn{1}{c}{Delay} & \multicolumn{1}{c}{Time} &       & \multicolumn{1}{c}{Area} & \multicolumn{1}{c}{Delay} & \multicolumn{1}{c}{Time} &       & \multicolumn{1}{c}{Area} & \multicolumn{1}{c}{Delay} & \multicolumn{1}{c}{Time} &       & \multicolumn{1}{c}{Area} & \multicolumn{1}{c}{Delay} & \multicolumn{1}{c}{Time} \\
\cmidrule{1-4}\cmidrule{6-8}\cmidrule{10-12}\cmidrule{14-16}\cmidrule{18-20}\cmidrule{22-24}    \multicolumn{1}{l}{Adder} & 790.73 & 1641.45 & 0.03  &       & 790.73 & 1641.45 & 0.06  &       & 815.72 & 1755.19 & 0.03  &       & \textbf{776.14} & \textbf{1199.59} & 0.31  &       & 941.08 & \textbf{875.02} & 0.2   &       & \textbf{662.94} & 2172.97 & 0.33 \\
    \multicolumn{1}{l}{Bar} & 1715.27 & 124.08 & 0.06  &       & 2015.64 & 119.56 & 0.13  &       & 1049.28 & 192.44 & 0.36  &       & \textbf{1706.64} & \textbf{113.82} & 1.21  &       & 2873.49 & \textbf{114.38} & 1.62  &       & 1088.16 & \textbf{187.74} & 0.99 \\
    \multicolumn{1}{l}{Div} & 14630 & 28460.6 & 0.4   &       & 15172.8 & 28503.4 & 1.69  &       & 12406.8 & 33895 & 2.67  &       & \textbf{13846.8} & \textbf{21337.3} & 9.18  &       & 17006.75 & \textbf{14816.8} & 23.89 &       & \textbf{12297.2} & \textbf{33317.1} & 11.55 \\
    \multicolumn{1}{l}{Hyp} & 171964 & 154106 & 2.08  &       & 158875 & 156469 & 321.67 &       & 149746 & 251234 & 285.81 &       & \textbf{163843} & \textbf{139545} & 12.48 &       & 165663.9 & \textbf{93686.7} & 51.95 &       & 153742 & \textbf{146165} & 15.71 \\
    \multicolumn{1}{l}{Log2} & 22093.4 & 2762.87 & 0.97  &       & 24377.2 & 2299.41 & 3.92  &       & 18471 & 4041.43 & 9.46  &       & \textbf{20660.5} & \textbf{2459.82} & 8.3   &       & 27347.54 & \textbf{2085.4} & 43.61 &       & \textbf{18358.3} & \textbf{3771.08} & 21.73 \\
    \multicolumn{1}{l}{Max} & 1477.73 & 1042.65 & 0.06  &       & 1852.63 & 1019.86 & 0.16  &       & 1170.16 & 1380.93 & 0.66  &       & \textbf{1346.4} & \textbf{996.11} & 0.66  &       & \textbf{1428.01} & \textbf{965.06} & 0.51  &       & \textbf{1082.09} & \textbf{1100.75} & 0.87 \\
    \multicolumn{1}{l}{Multiplier} & 20779 & 1845.75 & 0.43  &       & 20809.6 & 1841.55 & 1.45  &       & 16815.2 & 2883.92 & 3.44  &       & 21221.3 & \textbf{1475.42} & 3.6   &       & 28432.87 & \textbf{1313.29} & 5.3   &       & \textbf{15306.3} & \textbf{2553.4} & 7.93 \\
    \multicolumn{1}{l}{Sin} & 4500.62 & 1253.97 & 0.15  &       & 4860.48 & 1181.02 & 0.56  &       & 2919.4 & 2291.02 & 2.12  &       & 4622.06 & \textbf{1151.02} & 2.56  &       & 5889.58 & \textbf{1007.25} & 9.4   &       & 3073.9 & \textbf{1782.68} & 6.03 \\
    \multicolumn{1}{l}{Sqrt} & 19098.2 & 38212.1 & 0.24  &       & 18257.5 & 35005.6 & 3.98  &       & 10103.5 & 64636.5 & 5.46  &       & \textbf{16452} & \textbf{34349.3} & 1.25  &       & \textbf{13480.84} & \textbf{26358.8} & 1.48  &       & 11781.2 & \textbf{36746.8} & 1.66 \\
    \multicolumn{1}{l}{Square} & 12725.6 & 1556.2 & 0.22  &       & 12698.3 & 1558.31 & 0.7   &       & 12746.4 & 2180.82 & 1.38  &       & \textbf{12038.3} & \textbf{1206.85} & 1.57  &       & 12997.8 & \textbf{842.45} & 2.6   &       & \textbf{12131.5} & \textbf{1734.43} & 3.51 \\
    \multicolumn{1}{l}{Arbiter} & 1222.77 & 558.8 & 0.15  &       & 1276.27 & 562.04 & 2.48  &       & 948.2 & 598.87 & 4.96  &       & \textbf{1135.37} & \textbf{558.8} & 4.28  &       & 1436.4 & \textbf{553.81} & 5.21  &       & 986.5 & \textbf{581.54} & 2.31 \\
    \multicolumn{1}{l}{Cavlc} & 231.72 & 122.49 & 0.03  &       & 274.32 & 113.35 & 0.06  &       & 171.43 & 176.98 & 0.11  &       & 244.16 & \textbf{119.92} & 0.16  &       & \textbf{257.26} & 117.64 & 0.16  &       & 182.06 & \textbf{162.12} & 0.11 \\
    \multicolumn{1}{l}{Ctrl} & 56.46 & 64.78 & 0.03  &       & 56.23 & 64.78 & 0.04  &       & 40.81 & 117.8 & 0.02  &       & \textbf{55.09} & \textbf{64.1} & 0.11  &       & 62.52 & \textbf{62.61} & 0.1   &       & \textbf{40.58} & \textbf{97.8} & 0.06 \\
    \multicolumn{1}{l}{Dec} & 288.71 & 42.42 & 0.03  &       & 288.71 & 42.42 & 0.04  &       & 288.5 & 44.14 & 0.07  &       & 288.94 & 42.42 & 0.2   &       & 288.94 & 42.42 & 0.2   &       & \textbf{288.48} & \textbf{43.07} & 0.16 \\
    \multicolumn{1}{l}{I2c} & 525.32 & 114.36 & 0.04  &       & 553.24 & 101.3 & 0.07  &       & 466.12 & 290.33 & 0.13  &       & \textbf{485.44} & \textbf{113.1} & 0.41  &       & \textbf{534.05} & \textbf{95.96} & 0.48  &       & \textbf{426.28} & \textbf{192.75} & 0.39 \\
    \multicolumn{1}{l}{Int2float} & 86.26 & 115.65 & 0.03  &       & 104.88 & 89.33 & 0.04  &       & 78.04 & 151.16 & 0.03  &       & \textbf{85.83} & \textbf{114.36} & 0.12  &       & \textbf{91.21} & 105.23 & 0.11  &       & \textbf{77.18} & \textbf{140.22} & 0.07 \\
    \multicolumn{1}{l}{Mem\_ctrl} & 11672.5 & 666.61 & 0.6   &       & 12170.5 & 523.48 & 4.8   &       & 8779.48 & 1052.24 & 9.88  &       & \textbf{9285.38} & \textbf{630.69} & 9.89  &       & \textbf{10920.68} & \textbf{522.14} & 10.71 &       & 9482.35 & \textbf{1038.76} & 10.07 \\
    \multicolumn{1}{l}{Priority} & 237.65 & 433.19 & 0.04  &       & 235.81 & 433.19 & 0.06  &       & 226.9 & 424.58 & 0.1   &       & \textbf{225.97} & \textbf{424.58} & 0.47  &       & \textbf{222.27} & \textbf{421.38} & 0.46  &       & \textbf{223.58} & 453.28 & 0.35 \\
    \multicolumn{1}{l}{Router} & 91.86 & 129.6 & 0.02  &       & 103.08 & 119.5 & 0.04  &       & 78.8  & 180.49 & 0.06  &       & 108.7 & \textbf{116.77} & 0.16  &       & 112.87 & 112   & 0.19  &       & \textbf{75.99} & \textbf{151.22} & 0.12 \\
    \multicolumn{1}{l}{Voter} & 11442.8 & 492.94 & 0.13  &       & 11427 & 492.94 & 0.51  &       & 6850.8 & 831.15 & 0.88  &       & \textbf{10601.4} & \textbf{483.2} & 1.36  &       & 12771.23 & \textbf{485.14} & 3.42  &       & \textbf{6779.99} & \textbf{748.13} & 1.13 \\
    \midrule
    \multicolumn{1}{l}{Geomean} & 2039.15  & 813.86  & 0.11  &       & 2151.96  & 768.43  & 0.42  &       & 1623.93  & 1172.79  & 0.68  &       & \textbf{1963.02 } & \textbf{741.09 } & 1.06  &       & 2237.97  & \textbf{648.22 } & 1.53  &       & \textbf{1610.42 } & 1015.79  & 1.12  \\
    \multicolumn{1}{l}{Improvements} & -     & -     &       &       & -5.53\% & 5.58\% &       &       & 20.36\% & -44.10\% &       &       & \textbf{3.73\%} & \textbf{8.94\%} &       &       & -9.75\% & \textbf{20.35\%} &       &       & \textbf{21.02\%} & -24.81\% &  \\
    \bottomrule
    \end{tabular}%
  \label{ASIC mapping}%
  \vspace{-4mm}
\end{table*}%

\subsubsection{FPGA Technology Mapping}

The \emph{EPFL Combinational Benchmark Suite}'s \emph{Best Results Challenge} focuses on obtaining 6-LUT networks with the minimal LUT count. We use this challenge to evaluate the LUT mapping with MCH. Many studies have demonstrated its effectiveness through this challenge\footnote{\url{https://github.com/lsils/benchmarks/tree/master/best_results}}.

The experiment first converts the existing results into AIG networks using ABC's \texttt{strash} command. Note that the AIGs obtained through this conversion contain many redundant structures, and direct mapping of these AIGs yields poorer LUT netlists compared to the original results. We use the converted AIG networks as inputs to create the mixed choice networks. These mixed choice networks retain the original AIG structures while creating equivalent XMG candidate structures. Finally, we perform area-focused LUT mapping on the mixed choice networks. 

Table \ref{lutmapping} presents the results of the test cases in which MCH achieved new records. We set new records for the minimum 6-LUT count in four test cases compared to the \emph{best 6-LUT count results of 2023} (Best (2023)) and in three test cases compared to those of 2024 (Best (2024)). Notably, this experiment directly evaluated the effectiveness of the MCH-based mapper without any logic optimization or post-mapping optimization steps. 

In this experiment, our method dynamically evaluates the structural features of both XMG and AIG. MCH stores them as selectable substructures, rather than simplifying the entire circuit into a single logic representation for comparison. The results demonstrate that the combination of different logic representations effectively overcomes the limitations of a single logic representation structure.

\begin{table}[htbp]
  \centering
  \caption{Best area results for the EPFL benchmarks}
  \setlength{\tabcolsep}{1mm}
    \begin{tabular}{lrrrrrrrrr}
    \toprule
    \multicolumn{1}{c}{\multirow{2}[4]{*}{Benchmarks}} & \multicolumn{2}{c}{Best (2023)} & \multicolumn{2}{c}{MCH (2023)} & \multicolumn{2}{c}{Best (2024)} & \multicolumn{2}{c}{MCH (2024)} \\
\cmidrule{2-9}          & \multicolumn{1}{c}{Node} & \multicolumn{1}{c}{Lev} & \multicolumn{1}{c}{Node} & \multicolumn{1}{c}{Lev} & \multicolumn{1}{c}{Node} & \multicolumn{1}{c}{Lev} & \multicolumn{1}{c}{Node} & \multicolumn{1}{c}{Lev} \\
    \midrule
    Sin   & 1053  & 92    & \textbf{1051} & \textbf{81} & 1023  & 110   & \textbf{1020} & \textbf{107} \\
    Sqrt  & 2983  & 1526  & \textbf{2981} & \textbf{1374} & - & - & - & - \\
    Square & 2959  & 172   & \textbf{2956} & \textbf{167} & - & - & - & - \\
    Hyp   & - & - & - & - & 36507 & 4631  & \textbf{36506} & \textbf{4581} \\
    Voter & 1180  & 30    & \textbf{1179} & \textbf{28} & 1166  & 34    & \textbf{1165} & \textbf{27} \\
    \bottomrule
    \end{tabular}%
  \label{lutmapping}%
\end{table}%

\subsection{MCH for Logic Optimization}
This section conducts tests on the MCH-based logic optimization algorithm.
The experiment selected the graph mapping algorithm from~\cite{calvino2022versatile} (\emph{Graph Map}) for comparison. The experiment evaluated the logic optimization effect on the XMG network and the LUT mapping results (\emph{LUT Map}) of the optimized network. This experiment is conducted based on the mixed choice networks composed of MIG and XMG. The experiment performed iterative optimization on the logic network until no further improvement could be achieved, assessing the node count and logic level of the optimized XMG network. Subsequently, the optimized network was mapped to 6-LUT network, and the node count and logic level of the LUT network were evaluated.

\begin{figure}[htbp]
\centering
\includegraphics[scale=0.32]{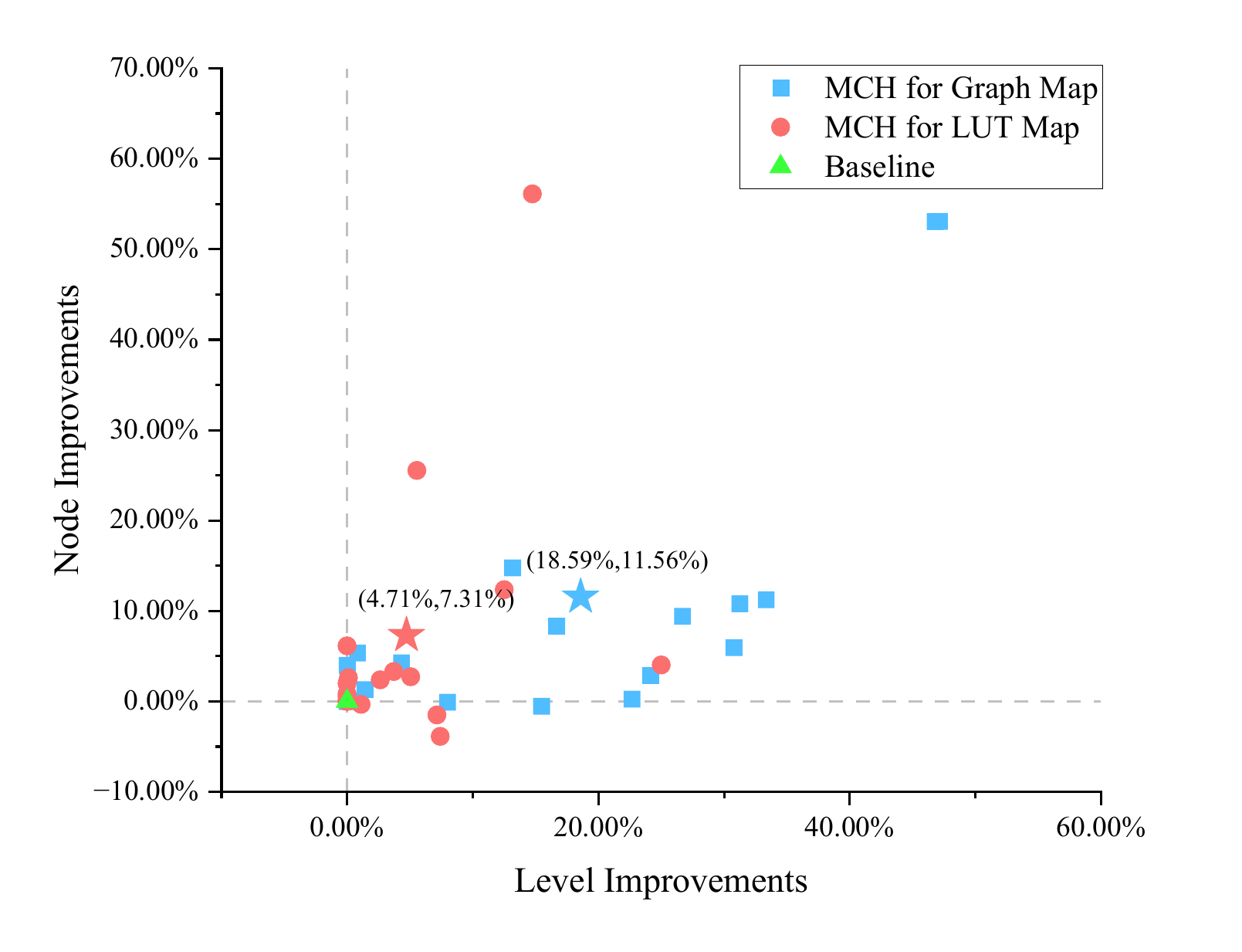}
\caption{Experimental results of Graph Map Optimization}
\label{Graph map}
\vspace{-4mm}
\end{figure}

The experimental results are presented in Fig.~\ref{Graph map}. Test cases also come from the \emph{EPFL Combinational Benchmark Suite}.
The experiments use the results of the original algorithm iterated to a local optimum as the baseline (Baseline). The horizontal and vertical axes represent the improvements in level and node count achieved by our method for the test cases. The blue squares in the figure represent the improvements in graph mapping optimization for each case under the influence of MCH (\emph{MCH for Graph Map}). The red circles represent the improvements for each case brought by the 6-LUT network generated from the optimized XMG using MCH-based LUT mapping (\emph{MCH for LUT Map}).
The blue and red star markers represent the geometric mean of MCH's gains for \emph{Graph Map} and \emph{LUT Map}, respectively. It is evident that with the assistance of MCH, most cases exhibit significant improvements. In some cases, after breaking through local optima, certain metrics improve by as much as 30\% or more. For some test cases, such as ``Round-robin arbiter" and ``Square-root", both the node count and level count were reduced by approximately 50\%. For \emph{Graph Map}, our method reduced the XMG level count by an average of 18.59\% and the node count by 11.56\%. After LUT mapping, the 6-LUT level count and node count were further reduced by 4.71\% and 7.31\%, respectively.

\section{Conclusion}
This paper presents a technology mapping framework based on mixed structural choices operator. The framework introduces mixed structural choices to retain multiple logic representations simultaneously and provides a rich set of candidates through an extensible multi-strategy structural choice method. MCH-based technology mapping effectively combines MCH with technology mapping algorithms, enabling dynamic evaluation of different logic representations and logic optimization candidates based on technology information. Additionally, the framework supports extension to various mapping-based optimization algorithms, helping to overcome the limitations of local optima. 
The method proposed in this paper is not only of reference value in heterogeneous synthesis and mapping of multi-logic domain, but also in  structural bias problems based on structure choices.

\clearpage
\footnotesize
\bibliographystyle{ieeetr}
\bibliography{ref}

\end{document}